\documentclass[preprint,showpacs,preprintnumbers,amsmath,amssymb]{revtex4-1}

\usepackage{graphicx}
\usepackage{dcolumn}
\usepackage{bm}
\usepackage{verbatim} 

\begin{document}
\title{Half-Chain Entanglement Entropy in the One-Dimensional Spinless Fermion Model}
\author{Myung-Hoon Chung\footnote{
To whom correspondence should be addressed.\\
$E$-$mail$ $address$: mhchung@hongik.ac.kr}}

\affiliation{College of Science and Technology, Hongik University, Sejong 339-701, Korea}

\date{\today}

\begin{abstract}
We calculate the half-chain entanglement entropy of
the ground state  in the one-dimensional spinless fermion model.
Considering a tiny corner of the Hilbert space represented by
matrix product states, we efficiently find the ground state by
the infinite time-evolving block decimation.
The Schmidt coefficients are used to determine the half-chain entanglement entropy.
Using the bond dimension scaling of the half-chain entanglement entropy,
we find the critical region, which is consistent with the previous results.
\end{abstract}


\maketitle

\section{Introduction}

In addition of quantum Monte Carlo \cite{Ceperley} and exact
diagonalization \cite{Dagotto}, the method of tensor networks
\cite{Orus} is one of the powerful theoretical tools to study
strongly correlated many-body systems. Tensor network algorithms
are based on the density-matrix renormalization group (DMRG)
\cite{White}. DMRG has proven to be a great success in the
simulation of strongly correlated one-dimensional quantum lattice
systems. The method of tensor network states has become popular
after we found that the internal structure of DMRG can be
understood with respect to the matrix product states (MPS)
\cite{Ostlund,Schollwoeck}. For two-dimensional systems, the
projected entangled-pair states (PEPS) \cite{Verstraete} are
introduced. More generally, there are many tensor network states
(TNS), which include MPS, PEPS, tree tensor network states
\cite{Murg}, the multiscale entanglement renormalization ansatz
(MERA) \cite{Vidal1}, matrix-product projected states \cite{Chou},
and projected entangled simplex states (PESS) \cite{Xie2}. These
tensor network states are used as the basis set for variational
approaches to quantum many-body systems.

When a total Hamiltonian is written as a sum of local
Hamiltonians, Vidal introduced a powerful method called
time-evolving block decimation (TEBD) \cite{Vidal2,Vidal3} to find
the ground state. If the total Hamiltonian has a symmetry such as
translational invariance, we can use the so-called infinite TEBD
(iTEBD) \cite{Vidal4}, in which we assume that the tensors in the
TNS have the same form, and we update a few tensors to achieve the
ground state.

Entanglement is a truly quantum mechanical phenomenon
\cite{Horodecki}. One of the recent interests in theoretical
condensed matter physics is to understand entanglement
\cite{Eisert} in strongly correlated quantum many-body systems. In
fact, the entanglement entropy is used to characterize quantum
phases \cite{Chen,Chung2}. To define the entanglement entropy, we
divide the quantum system into two parts $A$ and $B$. We introduce
a density matrix $\rho = |\Psi \rangle  \langle \Psi |$ with a
pure quantum state $|\Psi \rangle$ for the whole system, and
obtain the reduced density matrix of subsystem $A$ such as
$\rho_{A}=\mbox{Tr}_{B}\rho$ by tracing out subsystem $B$. The
entanglement entropy is the von Neumann entropy, which is given by
$S_{A} = -\mbox{Tr}(\rho_{A}\log_2 \rho_{A})$. In particular, for
$1+1$-dimensional quantum systems, the half-chain entanglement
entropy \cite{Tagliacozzo} is defined by using the Schmidt
coefficients. The entanglement entropy for a finite block of a
$1+1$-dimensional critical system was calculated analytically by
using conformal field theory \cite{Cardy}.

The purpose of this paper is to find the half-chain entanglement entropy
in the one-dimensional spinless fermion model.
To do so, we start by using the ansatz of matrix product state for the ground
state in the model. We present a slightly modified method in determining the ground state energy
by normalizing the maximum Schmidt coefficient \cite{Cha}. We apply our method to the
one-dimensional spinless fermion model. We also find the finite bond dimension scaling
of the half-chain entanglement entropy.
The scaling behavior is consistent with the formula of the entanglement entropy given by Calabrese and Cardy \cite{Cardy}.

This paper is organized as follows. In Sec. 2, a brief description
of our method is given, and we present the role of the maximum
Schmidt coefficient in determining the ground state energy. In
Sec. 3, by using iTEBD, we calculate the ground state for the
one-dimensional spinless fermion model. In Sec. 4, we present the
numerical results, which are the half-chain entanglement entropy,
and the corresponding data collapse. We find that the entanglement
entropy shows abrupt changes on the boundary of phases. In
conclusion, we discuss the future goal of simulating the
two-dimensional fermion model, where we encounter notorious
difficulties in relation with the negative sign problem.

\section{Method}

We briefly review the method \cite{Cha} that we use here for completeness.
There are two methods to obtain the ground state in tensor network algorithms.
One is the variational method, in which we minimize the energy
expectation value by changing parameters in the ansatz. The other is the imaginary time evolving, which
produces the ground state energy by applying an operator on
a tensor network state consecutively. The imaginary time evolving method
is adopted here to find the ground state.

The imaginary time evolving method starts by considering the
formal solution $|\Psi( T ) \rangle$ to the
imaginary time Schr\"{o}dinger equation, which is written as
\begin{equation}
|\Psi( T ) \rangle = \exp\{ - ( H - E ) T \} |\Psi( 0 )\rangle =  \prod^{T/\tau}\exp\{E\tau\}
\exp(-H\tau)|\Psi( 0 )\rangle  ,  \label{eq:dmcSolution}
\end{equation}
where we introduce an energy shift $E$ for a given Hamiltonian $H$, and
we will perform the Suzuki-Trotter decomposition for $\exp(-H\tau)$, where
the Trotter time step $\tau$ should be sufficiently small.
As the imaginary time $T$ goes to infinity, the state $|\Psi( T )
\rangle$ becomes the ground state for properly chosen $E$. In
fact, when $E$ is larger or smaller than the ground-state energy,
$|\Psi( T )\rangle$ blows up or shrinks down, respectively, in the
limit as $T\rightarrow \infty$. In a numerical approach, we
redefine $E$ as a function of $T$ to maintain the norm of state.

For one-dimensional systems, our ansatz for the ground state
is a MPS because of the area law \cite{Eisert}.
A tensor in MPS, $A_{ab}^{\sigma}$, has three indices, among which the physical
index $\sigma$ takes a value from $0$ to $d-1$. For the internal
bond degree of freedom, the indices $a$ (left) and $b$ (right) run
from $0$ to $\chi-1$, where $\chi$ is the bond dimension. The
Schmidt coefficients between $A_{ab}^{\sigma}$ and $B_{bc}^{\rho}$
are denoted by $\lambda^{AB}_{b}$. A state in the space of the
matrix product states is written as
\begin{equation}
|\mbox{MPS} \rangle = \sum_{\cdots \sigma\rho\nu\eta\cdots}
\mbox{Tr}(\cdots \lambda^{XA}_{a} A^{\sigma}_{ab} \lambda^{AB}_{b} B^{\rho}_{bc}
\lambda^{BC}_{c} C^{\nu}_{cd} \lambda^{CD}_{d} D^{\eta}_{de}
\cdots )| \cdots \sigma\rho\nu\eta\cdots \rangle  ,   \label{eq:MPS}
\end{equation}
where $\mbox{Tr}$ means that all internal bond indices $a, b, c,
d, \cdots$ are summed up.

Vidal proposed a clever idea of TEBD for updating the tensors in
the MPS of Eq. (\ref{eq:MPS}). Usually a local Hamiltonian is written as a
sum of the elementary operators $h_{ij}$ such as $H=\sum_{ij}h_{ij}$, where the index of $i$
and $j$ represents nearby sites. Then, by the Suzuki-Trotter decomposition, it is enough to
consider the effect of $\exp(-h_{ij}\tau)$.
When $\exp(-h_{ij}\tau)$ acts on the previous MPS, we approximate the output state
into our new MPS by updating the tensors and the Schmidt
coefficients locally. In fact, the first step in TEBD is to find the
four-index tensor $M$ defined by
\begin{equation}
M^{\rho_{i}\rho_{j}}_{\sigma_{i} \sigma_{j}} = \langle
\rho_{i}\rho_{j} | \exp(-h_{ij}\tau) | \sigma_{i} \sigma_{j}
\rangle  .
\end{equation}
For example, in order to update $A$, $B$, and $\lambda^{AB}$ in
$|\mbox{MPS} \rangle$ of Eq. (\ref{eq:MPS}), we consider the four-index tensor $\Theta$:
\begin{equation}
\Theta^{\rho_{i}\rho_{j}}_{ac} = \sum_{b=0}^{\chi -1}
\sum_{\sigma_{i},\sigma_{j}=0}^{d -1} \lambda^{XA}_{a}
A^{\sigma_{i}}_{ab} \lambda^{AB}_b B^{\sigma_{j}}_{bc}
\lambda^{BC}_{c} M^{\rho_{i}\rho_{j}}_{\sigma_{i} \sigma_{j}}.
\end{equation}
By employing singular value decompositions (SVD), we obtain the
updated $\tilde{\lambda}^{AB}_{b}$ by keeping the $\chi$ largest
weights:
\begin{equation}
\Theta^{\rho_{i}\rho_{j}}_{ac} ~\rightarrow~ \sum_{b=0}^{\chi-1}
\bar{A}^{\rho_{i}}_{ab} \tilde{\lambda}^{AB}_b
\bar{B}^{\rho_{j}}_{bc}=\sum_{b=0}^{\chi-1}
\lambda^{XA}_{a}\tilde{A}^{\rho_{i}}_{ab} \tilde{\lambda}^{AB}_b
\tilde{B}^{\rho_{j}}_{bc}\lambda^{BC}_{c}.
\end{equation}
By dividing and attaching the weights, we find $\tilde{A}$ and
$\tilde{B}$ from $\bar{A}$ and $\bar{B}$ in the above.
We denote this process graphically as follows:
\begin{equation*}
\begin{array}{ccccccccccc}
-& &\Theta  & &-&     \rightarrow   &-&\tilde{A}&-&\tilde{B}&- \\
 &|&        &|& &                   & &        |& &        |&
\end{array}
\end{equation*}
where the vertical lines mean the physical indices. A similar
procedure is performed for other tensors $C$, $D$, $\cdots$ and
other weights on the bonds. We note that it is possible to update
the tensors simultaneously if we use multi-core computers. If
there is a translational symmetry in the local Hamiltonian, we use
iTEBD, where we assume that only two matrices $A$ and $B$ repeat
in the MPS. In this case, we need to update $A$, $B$,
$\lambda^{AB}$, and $\lambda^{BA}$ only.

Now we present how to determine the ground state energy without
evaluations of inner products \cite{Chung0}. Being inspired by the diffusion
Monte Carlo \cite{Ceperley,Chung1}, we have introduced the energy $E$
in front of the operator of the Suzuki-Trotter decomposition in Eq. (\ref{eq:dmcSolution}).
While the energy is adjusted by
controlling the number of replicas in the diffusion Monte Carlo, here we determine $E$ by
managing the factor in front of the wave function \cite{Chung0}.
The algorithm is as follows: when a typical operator
$\exp(-h_{ij}\tau)$ acts on a matrix product state $|\mbox{MPS}
\rangle$, we perform SVD and obtain $\chi$ singular values of
$\lambda_{0} \ge \lambda_{1} \ge \cdots \ge \lambda_{\chi -1}$. We
take out the maximum $\lambda_{0}$ and place it in front of the wave function, and
we modify the singular values as follows: $1 \ge
\lambda_{1}/\lambda_{0} \ge \cdots \ge \lambda_{\chi
-1}/\lambda_{0}$. In this way, we normalize $|\mbox{MPS} \rangle$
such that all weights on each bond have the maximum value of 1.
Thus, whenever the weights are modified by $\exp(-H\tau)$ acting
on the $k$-th time step state $|\mbox{MPS}_{k} \rangle$, we take
out the maximum weight to obtain the factor $F$ in front of the state
\begin{equation}
\exp(- H\tau)|\mbox{MPS}_{k} \rangle = F|\mbox{MPS}_{k+1} \rangle ,
\end{equation}
where the maximum Schmidt coefficient on each bond in
$|\mbox{MPS}_{k+1} \rangle$ is equal to $1$. We obtain the factor
$F$ by multiplying the previous $F$ by $\lambda_{0}$ whenever any bond is modified.
Because we require no divergence and no convergence to zero for
the state, as in the diffusion Monte Carlo, we adjust the next
energy value $E_{k+1}$ for the state to be stable in this way:
\begin{equation}
E_{k+1} = - \frac{\log F}{\tau}. \label{eq:dE}
\end{equation}
After we find $E_{k+1}$, we set $F=1$ again for the next iteration
in the computer simulation. We note that during the time
evolution, $E_{k}$ is stable and approaches the ground-state
energy in the limit of $k \rightarrow \infty$. The solution of
$|\mbox{MPS}_{\infty} \rangle$ is also stable.

\section{One-dimensional spinless fermion model}

In relation with the Luttinger liquids, the spinless fermion model may be the simplest fermion system, which
is a good laboratory for a benchmark calculation \cite{Chung0}.
We begin by presenting the
Hamiltonian for the one-dimensional spinless fermion model:
\begin{equation}
H = -t\sum_{i}(c^{\dagger}_{i}c_{i+1} +
c^{\dagger}_{i+1}c_{i}) + V\sum_{i}(n_{i}-\frac{1}{2})(n_{i+1}
-\frac{1}{2}) -\mu \sum_{i}n_{i}, \label{eq:H}
\end{equation}
where $t$ is the nearest-neighbor hopping strength in
a one-dimensional lattice, $V$ is the nearby Coulomb interaction strength,
and $n_{i}$ is the number operator.
The role of the chemical potential $\mu$ is to control the number
of fermions in the system.
Because this one-dimensional spinless fermion model
preserves the translational invariance, we can use iTEBD here.

When we apply iTEBD to find the ground state, we divide the Hamiltonian
into two parts, which are denoted by $e$(even) and $o$(odd), such
as $H = H_{e} + H_{o}=\sum_{j}h_{2j~2j+1}+\sum_{j}h_{2j+1~2j+2}$ for the Suzuki-Trotter decomposition.
The elementary operators $h_{i~i+1}$ are written as
\begin{eqnarray}
h_{i~i+1} = -t(c^{\dagger}_{i}c_{i+1} +
c^{\dagger}_{i+1}c_{i}) + V(n_{i}-\frac{1}{2})(n_{i+1}
-\frac{1}{2}) -\frac{\mu}{2}n_{i} -\frac{\mu}{2}n_{i+1}.
\end{eqnarray}
In order to find the ground state, we consider a tiny corner of the Hilbert space. The tiny
corner is characterized by the MPS with two tensors and two vectors with a
fixed bond dimension $\chi$.

It is natural to define the physical index $\sigma_{i}$ in the MPS
for the spinless fermion model. The state on the $i$-th site is
represented by $\sigma_{i}=0$ or $1$, such as
$0$ for the vacancy and $1$ for the occupancy. The
state of the Fock space is written in terms of the creation
operators $c^{\dagger}_{i}$ as follows:
\begin{equation}
| \sigma_{0} \cdots \sigma_{L-1} \rangle =
(c^{\dagger}_{0})^{\sigma_{0}} \cdots
(c^{\dagger}_{L-1})^{\sigma_{L-1}} |0 \rangle  .
\end{equation}

Now we apply iTEBD in order to obtain the ground state of the
Hamiltonian of Eq. (\ref{eq:H}). In the process of iTEBD, we need
to determine $\langle \rho_{i}\rho_{i+1} | \exp (-h_{i~i+1}\tau) |
\sigma_{i} \sigma_{i+1} \rangle$. We first calculate the $4 \times
4$ matrix $\langle \rho_{i}\rho_{i+1} | (-h_{i~i+1}) | \sigma_{i}
\sigma_{i+1} \rangle $, which is written as
\begin{equation*}
\left( \begin{array}{cccc}
-\frac{1}{4}V & 0                               & 0                                & 0  \\
0             & \frac{1}{4}V+\frac{1}{2}\mu     & t                                & 0  \\
0             & t                               & \frac{1}{4}V+\frac{1}{2}\mu      & 0 \\
0             & 0                               & 0                                & -\frac{1}{4}V+\mu
\end{array} \right)
\end{equation*}
By using the Taylor expansion, the four-index tensor $M$ of $\langle \rho_{i}\rho_{i+1} | \exp (-h_{i~i+1}\tau) |
\sigma_{i} \sigma_{i+1} \rangle$ is written in terms of $\langle
\rho_{i}\rho_{i+1} | (-h_{i~i+1}\tau) | \sigma_{i} \sigma_{i+1} \rangle $. We obtain the values of $M$
numerically for given $t$, $V$, $\mu$, and $\tau$. For a
small $\tau$, it is enough to make the Taylor expansion up to ${\cal O}(
\tau^{7})$.

Performing iTEBD with the MPS, we use two matrices $A$, $B$ and two
Schmidt coefficients $\lambda^{AB}$, $\lambda^{BA}$.
Our goal is to find $|\mbox{MPS} \rangle$ for the ground state by
optimizing $A$ and $B$ in iTEBD. We describe the
procedure for the computational simulation of iTEBD from the
Suzuki-Trotter decomposition with $H = H_{e} + H_{o}$:

\begin{enumerate}
\item Choose $A$, $B$, $\lambda^{AB}$, and $\lambda^{BA}$ randomly.

\item Update $A$, $\lambda^{AB}$, and $B$, by handling $H_{e}$.

\item Update $B$, $\lambda^{BA}$, and $A$, by handling $H_{o}$.

\item Repeat from step 2 until  $\lambda^{AB}$ and $\lambda^{BA}$ are identical.
\end{enumerate}

\begin{figure*}
\includegraphics[width= 12.0 cm]{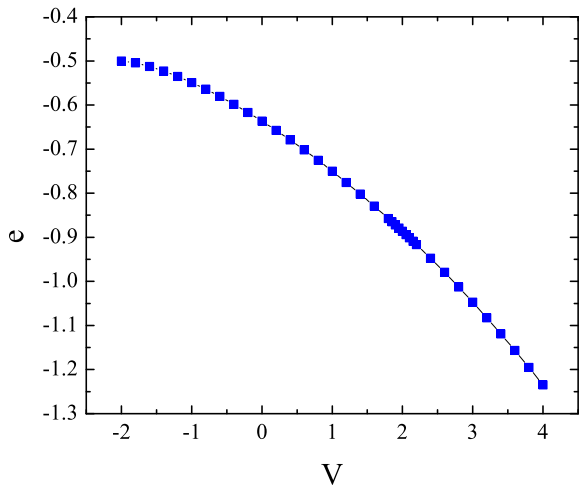}
\caption {The ground state energy per site $e$ versus $V$. We observe that
the energy values are almost identical for sufficiently large $\chi$.
We find the smooth behavior near $V=2$. We fix the
parameters as $t=1$, $\mu=0$, and $\tau=0.02$. }
\label{fig:gse}
\end{figure*}

From Eq. (\ref{eq:dE}), we note that the
corresponding energy per site $e$ is given by
\begin{equation}
e = - \frac{\log(\lambda^{AB}_{0}\lambda^{BA}_{0})}{2\tau},
\end{equation}
where $\lambda^{AB}_{0}$ and $\lambda^{BA}_{0}$ are the maximum
singular values obtained by performing SVD.
As a benchmark calculation, we present the ground state energy per site
for the model as shown in Fig \ref{fig:gse}.

\section{Numerical Results: Half-Chain Entanglement Entropy}

\begin{figure*}
\includegraphics[width= 12.0 cm]{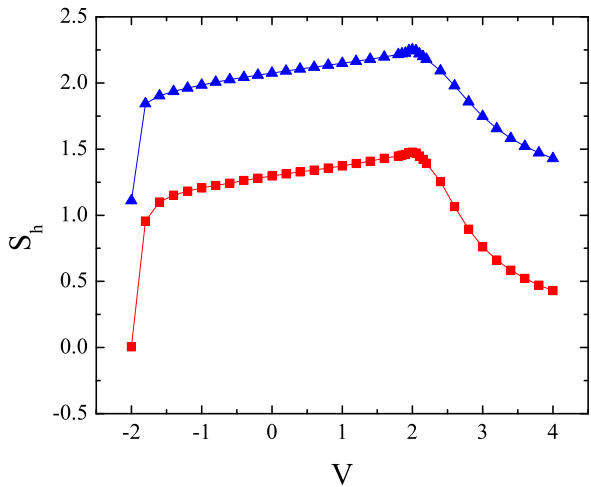}
\caption {The two kinds of the half-chain entanglement entropy $S_h$ versus $V$ for
$\chi = 28$. We fix the parameters as $t=1$, $\mu=0$, and $\tau=0.02$.
We note that simulation is unstable near $V=-2$ because of strong attraction between fermions.}
\label{fig:Sh28}
\end{figure*}

When we find the ground state as a form of MPS, the advantage of MPS
is easily to extract the half-chain entanglement entropy \cite{Tagliacozzo}.
We present the von Neumann entropy for the half-infinite chain by using the Schmidt coefficients:
\begin{equation}
S_{h} = - \sum_{i=0}^{\chi - 1} \lambda_{i}^{2} \log \lambda_{i}^{2},
~~~~~\lambda_{i} =
\frac{\lambda_{i}^{AB}}{\sqrt{\sum_{i}|\lambda_{i}^{AB}|^{2}}}.
\label{eq:hcee}
\end{equation}
We note that we do not have to distinguish $\lambda_{i}^{BA}$ from
$\lambda_{i}^{AB}$ in $|\mbox{MPS}_{\infty} \rangle$ because we
simulate until they become equal. By simulation with randomly
chosen initial states for the model, we repeatedly find two ground
states whose energy difference is invisible, in fact, less than
$10^{-5}$. Thus, it is hard to distinguish two states by energy.
However, two states prominently have different entanglement
entropies. We show two kinds of the corresponding half-chain
entanglement entropy $S_{h}$ in Fig. \ref{fig:Sh28} for the bond
dimension $\chi=28$. The shapes of entropy are similar, but the
values are quite different from each other.

\begin{figure*}
\includegraphics[width= 12.0 cm]{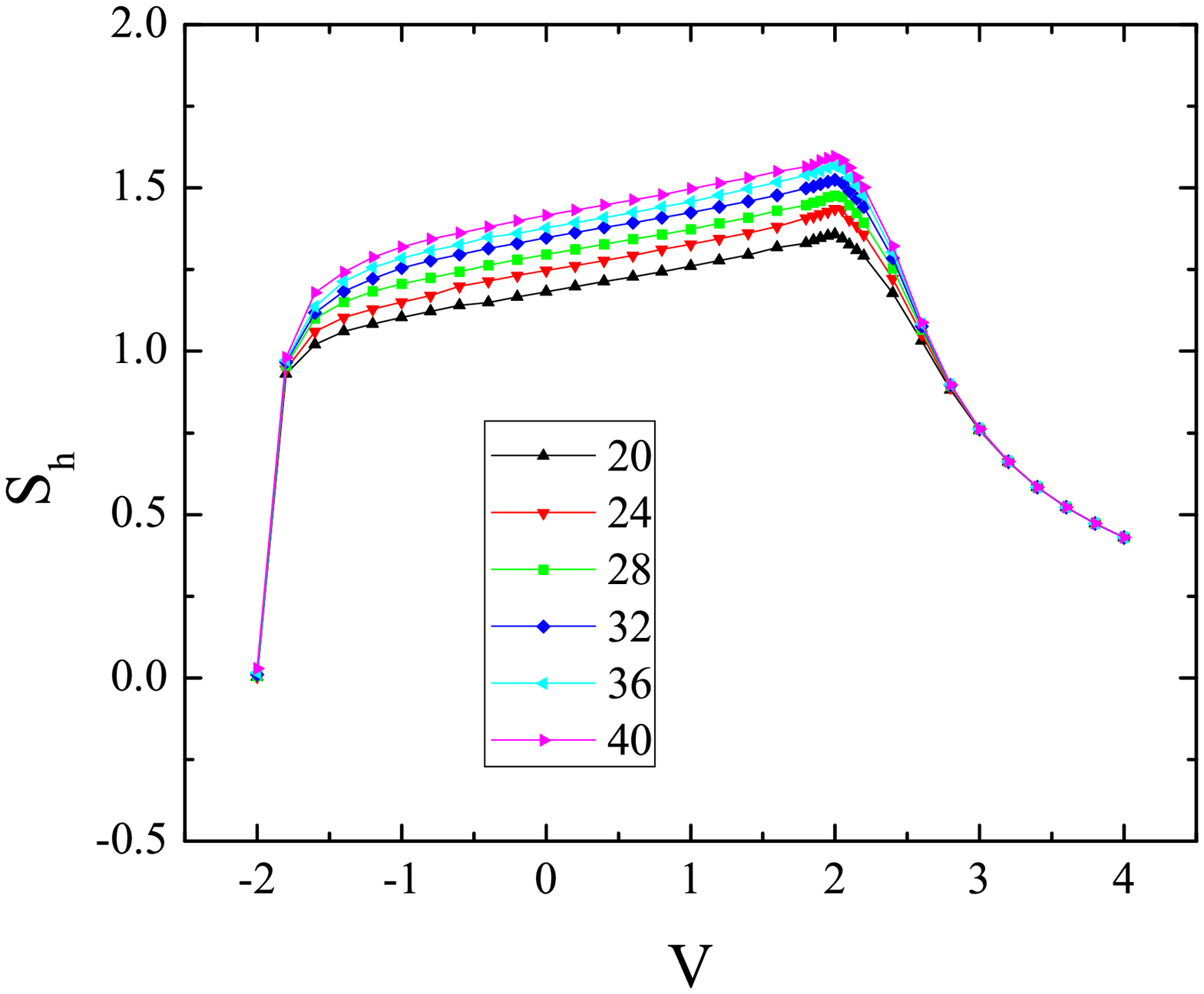}
\caption {The half-chain entanglement entropy $S_h$ versus $V$ for
several $\chi$ from $20$ to $40$. We fix the parameters as $t=1$, $\mu=0$, and $\tau=0.02$.}
\label{fig:Sh}
\end{figure*}

Focusing on the less entangled state which is lower in Fig.
\ref{fig:Sh28}, we present the half-chain entanglement entropy for
several internal bond dimensions $\chi$ in Fig. \ref{fig:Sh}. We
find the abrupt changes \cite{Chung0} near $V = 2$ and $V=-2$. For
$V > 2$, the entanglement entropy becomes saturated, in other
words, the entropy does not depend on $\chi$. This suggests that
the corresponding Hamiltonian is gapful. On the other hand, for
$-2 < V < 2$, as $\chi$ increases, the entropy also increases.
This $\chi$-dependence is the signal of criticality or gapless.

It is remarkable that the entanglement entropy for a critical system
can be obtained by means of conformal field theory. For the case where the whole
one-dimensional system has a finite but large length $L$ with the periodic
boundary condition and the subsystem is a single interval of
length $l$, the entanglement entropy \cite{Cardy} is given by
\begin{equation}
S_{l} = \frac{c}{3}\log(\frac{L}{\pi a}\sin(\frac{\pi l}{L})) +
c_{1}, \label{eq:Cardy}
\end{equation}
where $a$ is the lattice spacing, $c_{1}$ is a
non-universal constant, and $c$ is the central charge.

\begin{figure*}
\includegraphics[width= 12.0 cm]{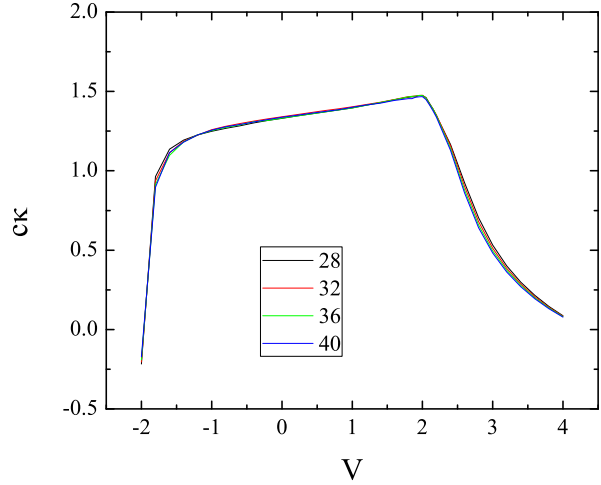}
\caption {The plot of $6(S_{h}-c_{1}^{\prime\prime})/\log(\chi)$
versus $V$ with $\chi=28,~32,~36,~40$ for the system of $t=1$,
$\mu=0$, and $\tau = 0.02$. We fix the non-universal
$c_{1}^{\prime\prime}$ as $c_{1}^{\prime\prime}=0.23+0.025V$.}
\label{fig:ES}
\end{figure*}

Because there is only one side contact for the half-chain,
we should modify the above formula for
the half-chain entanglement entropy by dividing 2
and setting $l=L/2$ such as
\begin{equation}
S_{h} = \frac{c}{6} \log (L)+c_{1}^{\prime}, \label{eq:EEh}
\end{equation}
where $c_{1}^{\prime}$ is a non-universal constant.
The length of the system size $L$ is related to the correlation length $\xi$
near a critical point such as $L\sim \xi$. Furthermore, it is postulated that the correlation length $\xi$ has the scaling
relation with the bond dimension $\chi$ such as $\xi \sim \chi^{\kappa}$ \cite{Tagliacozzo, Pollmann}.
Then, we find
\begin{equation}
S_{h} (\chi) = \frac{c\kappa}{6} \log (\chi)+c_{1}^{\prime\prime}. \label{eq:EEf}
\end{equation}
Using this formula, it is of interest to check the finite $\chi$
scaling of $S_{h}$. We find the data collapse in the plot of
$6(S_{h}-c_{1}^{\prime\prime})/\log(\chi)$ versus $V$ by adjusting
$c_{1}^{\prime\prime}$ as shown in Fig \ref{fig:ES}. The data
collapse happens between $V=-2$ and $V=2$, which is consistent
with the previous results about the critical region. Finally, we
determine the scaling exponent $\kappa$ from the data collapse of
Fig \ref{fig:ES}, for instance, $\kappa = 1.47(1)$ at $V=2$ with
the well-known result of $c=1$ \cite{Nishimoto}.

\section{Conclusion}

In summary, we have presented the half-chain entanglement entropy
for the spinless fermion model. We perform the finite bond dimension scaling
with the entanglement entropy. We find the critical behavior between $V=-2$ and $V=2$,
which is consistent with the previous results.

When we use multi-core computers, it is possible to parallelize
the local updates of the internal bonds for the model of no
translational symmetry. As a future work, we hope that someone will perform this
parallel computing soon.

It is of interest to extend our method to two-dimensional systems.
Although we do not encounter the notorious sign problem here,
we have to overcome the sign problem
in the two-dimensional model with PEPS. We anticipate
progress in the two-dimensional case.

\section*{Acknowledgments}
This work was partially supported by the Basic Science Research
Program through the National Research Foundation of Korea (NRF)
funded by the Ministry of Education, Science and Technology (Grant
No. 2011-0023395). The author would like to thank M.C. Cha and J.W. Lee for
helpful discussions.

\end{document}